\documentclass[aps,prb,reprint,groupedaddress,showpacs]{revtex4-1}

\usepackage{graphicx}
\usepackage{dcolumn}
\usepackage{bm}

\makeatletter
\AtBeginDocument{\@ifpackageloaded{natbib}{\ifNAT@numbers\if@filesw\immediate\write\@auxout{\string\global\string\NAT@numberstrue}\fi\fi}{}}
\makeatother
\usepackage{amsmath}

\begin{document}

\title{Observation of finite excess noise in the voltage-biased quantum Hall regime as a precursor for breakdown}

\author{Kensaku Chida,$^1$}\email{chida@scl.kyoto-u.ac.jp}
\author{Tomonori Arakawa,$^1$ Sadashige Matsuo,$^1$ Yoshitaka Nishihara,$^1$ Takahiro Tanaka,$^1$ Daichi Chiba,$^1$ Teruo Ono$^1$ Tokuro Hata,$^2$ Kensuke Kobayashi,$^{1, 2}$}
\author{Tomoki Machida$^{3, 4,}$}

\affiliation{$^1$Institute for Chemical Research, Kyoto University, Uji, Kyoto 611-0011, Japan}
\affiliation{$^2$Graduate School of Science, Osaka University, 1-1 Machikaneyama, Toyonaka, Osaka 560-0043, Japan.}
\affiliation{$^3$Institute of Industrial Science, University of Tokyo, 4-6-1 Komaba, Meguro-ku, Tokyo 153-8505, Japan}
\affiliation{$^4$Institute for Nano Quantum Information Electronics, University of Tokyo, 4-6-1 Komaba, Meguro-ku, Tokyo 153-8505, Japan}

\begin{abstract}
We performed  noise measurements in a two-dimensional electron gas to investigate the nonequilibrium quantum Hall effect (QHE) state.
While excess noise is perfectly suppressed around the zero-biased QHE state reflecting the dissipationless electron transport of the QHE state, considerable finite excess noise is observed in the breakdown regime of the QHE.
The noise temperature deduced from the excess noise is found to be of the same order as the energy gap between the highest occupied Landau level and the lowest empty one.
Moreover, unexpected finite excess noise is observed at a finite source-drain bias voltage smaller than the onset voltage of the QHE breakdown, which indicates finite dissipation in the QHE state and may be related to the prebreakdown of the QHE.
\end{abstract}

\date{\today}


\maketitle
\section{Introduction}
A two-dimensional electron gas (2DEG) under the influence of a large perpendicular magnetic field exhibits a remarkable dissipationless state with precisely quantized Hall resistance, which is the integer quantum Hall effect (QHE), \cite{Klitzing1980PRL} as a consequence of the topological rigidity of the phase. \cite{Thouless1982PRL} The existence of the localized bulk states plays an essential role in the precise quantization of the Hall resistance. \cite{Halperin1982PRB}
They spatially separate the counterflowing channels at the sample edge to strongly suppress  backscattering. \cite{Buttiker1988PRB}
More precisely, the edge channels are regarded as ``incompressible strips" owing to the electron screening. \cite{Chklovskii1992PRB, Lier1994PRB, Weis2012PTRSA}
Although the nature of the QHE state is successfully explained by the topological rigidity, details of the state in nonequilibrium are largely unexplored.
Recent experiments aimed at quantum information processing by using edge channels, involving phase reversal of electrons in electron interferometers, \cite{Neder2006PRL, Yamauchi2009PRB} decoherence, \cite{Litvin2007PRB, Roulleau2008PRL, Litvin2008PRB, Levkivskyi2008PRB, Youn2008PRL, McClure2009PRL} energy relaxation, \cite{ Altimiras2010Nphys, LeSueur2010PRL, Altimiras2010PRL} and dynamics of the edge magnetoplasmons in the edge channels, \cite{Kamata2010PRB, Kumada2011PRB} have clarified electron behaviors in the nonequilibrium QHE state.
Thus, detailing the nonequilibrium properties of the QHE states has become an important research topic in present condensed-matter physics.

The longstanding problem of the nonequilibrium QHE state called the QHE breakdown has vexed metrology researchers for about three decades.
The quantized Hall resistance  collapses at finite source drain voltage $V_{\rm sd}$ and/or at a current larger than a certain value. \cite{Cage1983PRL, Ebert1983JPC}
The QHE breakdown is, in other words, a transition from a topologically protected phase to a completely different one.
For this reason, the importance of the QHE breakdown has invoked renewed interest as new types of topologically protected phases, \cite{Hasan2010RMP} that is, topological insulators, have gathered much attention today.
As the topological insulator phase is a scion of the QHE state, a detailed understanding of how the QHE state breaks down should shed new light on the robustness of the topologically protected phases.

A number of mechanisms for the QHE breakdown have been proposed thus far \cite{Tsui1983BAPS, Trugman1983PRB, Streda1984JPC, Heinonen1984PRB, Eaves1986SST, Dyakonov1991SSC, Chaubet1995PRB, Chaubet1998PRB, Tsemekhman1997PRB}, and results of experiments have revealed substantial properties of the breakdown, including electron overheating through the avalanche-type electron scattering, \cite{Komiyama1985SSC, Komiyama2000PRB} the spatial gradient of the effective electron temperature along the electron path \cite{Komiyama1996PRL}, the typical length of the electron heating and cooling, \cite{Kaya1998PRB, Kaya1999EPL} and the time scale of the breakdown. \cite{Sagol2002PRB, Kawaguchi1995JJAP, Buss2005PRB}
Although these experiments have uncovered important information regarding the QHE breakdown, its mechanism remains to be clarified. \cite{Nachtwei1999PE}

Noise measurement is a promising candidate that can provide highly significant information on the QHE breakdown.
In fact, it has revealed various unique properties of electron transport \cite{Blanter2000PR} that could not be obtained through conventional conductance and resistance measurements.
In the QHE state, the excess noise is suppressed because of the absence of  backscattering. \cite{Buttiker1990PRL}
Hence, we expect that we can detect the QHE breakdown by investigating the mechanism through which excess noise occurs.
The excess noise may provide us with direct information on the effective electron temperature, which can be compared to the values calculated from the longitudinal resistance. \cite{Komiyama1985SSC, Komiyama2000PRB}
To the best of our knowledge, there has been no experimental work that has studied these aspects.

In this work, we present the experimental results of the noise measurement to clarify the QHE breakdown mechanism.
Two distinct observations are made. 
The first one is that the electron heating accompanied by the breakdown is of the order of the energy gap between the highest occupied Landau level and the lowest unoccupied one. 
Second, unexpectedly, we observed finite excess noise prior to the QHE breakdown at a finite $V_{\rm sd}$ that is smaller than the onset voltage of the QHE breakdown (a $V_{\rm sd}$ region that we define as a precursor regime), indicating the presence of finite dissipation in the nonequilibrium QHE states.
The excess noise behaves in a way closely related to the prebreakdown of the QHE \cite{Ebert1983JPC, Komiyama1985SSC, Meziani2004JAP}. 
Thus, the noise measurement may give us profound information about onset of the breakdown of the QHE.

This paper is organized as follows.
In Sec. II, we describe our device and the experimental setup used for high-frequency (2.55 MHz) and low-frequency (1--100 kHz) noise measurements.
In Sec. III A, we present the experimental results of the noise measurements, and using the results, we deduce the effective electron temperature in the breakdown regime of the QHE.
In Sec. III B, we show finite excess noise in the precursor regime. 
In Sec. IV, we summarize our work. 

\section{Experiment}
\subsection{Device}
\begin{figure}[tb] \center \includegraphics[width=.7\linewidth]{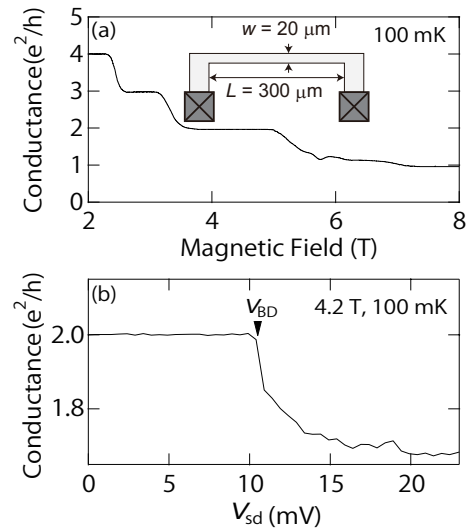}\caption{(a) Equilibrium conductance of the device as a function of $B$ and schematic illustration of the present two-terminal device. (b) Differential conductance of the device as a function of $V_{\rm sd}$. $V_{\rm BD}$ is defined as $V_{\rm sd}$ at which $\Delta G = G(V_{\rm sd}) - G(0) = 0.01$$e^2/h$.}\label{fig:dev}\end{figure}

The QHE breakdown was investigated using a two-terminal device fabricated on a semiconductor with a 2DEG in an AlGaAs/GaAs interface.
The 2DEG has electron density $n_e = 2.3 \times 10^{15}$ m$^{-2}$ and mobility $\mu = 110$ m$^2$/(V\(\,\)s).
The geometry of the two-terminal device is schematically shown in Fig. \ref{fig:dev}(a).
The Hall bar with  width $W =$ 40 $\mu$m is fabricated by  wet etching.
Then, the main part of the device is defined by the negatively charged gate electrodes to have $W =$ 20 $\mu$m and  length $L =$ 300 $\mu$m.
As Kaya \textit{et al.} reported previously, \cite{Kaya1998PRB, Kaya1999EPL} the narrow constriction of the main part is utilized for inducing the QHE breakdown specifically at the entrance of the main part.

The basic properties of the device were checked by a conductance measurement at  equilibrium with the standard lock-in technique with excitation voltage of 10 $\mu$V at 37 Hz.
Figure \ref{fig:dev}(a) shows the differential conductance of the device $G(V_{\rm sd})$ as a function of a  magnetic field perpendicular to the 2DEG $B$ at  electron temperature $T = 100$ mK.
The clear conductance plateaus at $e^2/h$, $2e^2/h$, $3e^2/h,$ and $4e^2/h$ represent the QHE with Landau level filling factors of $\nu = 1$, 2, 3, and 4, respectively.
At the conductance plateaus, the conductance remains quantized up to the finite source-drain bias voltage $V_{\rm sd}$ smaller than a certain critical value.
Figure \ref{fig:dev}(b) shows $G(V_{\rm sd})$ as a function of $V_{\rm sd}$ at $B = 4.2$ T ($\nu = 2.1$); the curve exhibits an abrupt conductance collapse from $2e^2/h$ at $V_{\rm sd} = 10.4$ mV.
In this paper, the onset voltage of the QHE breakdown $V_{\rm BD}$ is defined as the $V_{\rm sd}$ value at which the conductance deviation $\Delta G(V_{\rm sd}) = G(0) - G(V_{\rm sd})$ exceeds $0.01 e^2/h$. We call the $V_{\rm sd}$ region $|V_{\rm sd}| < |V_{\rm BD}|$ the ``QHE regime" and call the $|V_{\rm sd}| > |V_{\rm BD}|$ region the ``breakdown regime."
We focus on the results obtained with the etching-defined edge as it exhibits a more distinctive QHE breakdown than the electrically defined one. \cite{asymmetric}

\subsection{Noise measurement and its analysis}
To obtain the noise spectral density of the device, a high-frequency (approximately megahertz) noise measurement setup and a low-frequency (approximately kilohertz) noise measurement setup were utilized.
The former adopts a resonator to obtain information at a frequency that is sufficiently high to prevent $1/f$ noise by focusing on a specific frequency defined by the resonator. 
The latter  enables us to obtain the frequency dependence of the noise spectral density, although the bandwidth is limited to less than 100 kHz owing to  $RC$ damping.
From the noise spectrum for a wide frequency range below 100 kHz, the contribution of low-frequency noise such as $1/f$ noise is evaluated.
In the subsequent text, we describe the measurement scheme and analysis in further detail.

\subsubsection{High-frequency measurements}
\begin{figure}[b]\center \includegraphics[width=1.0\linewidth]{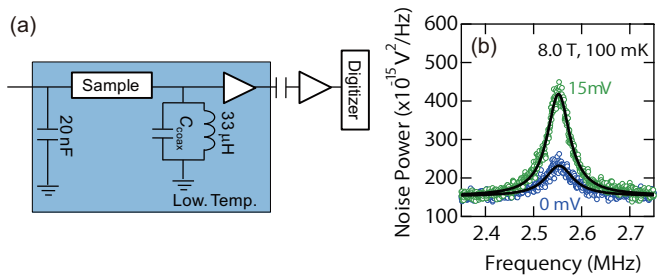}\caption{(Color online) (a) Schematic illustration of the two-terminal device and setup for the high-frequency noise measurements. (b) Noise spectra with a resonant peak at 2.55 MHz obtained at $V_{\rm sd} = 0$  and 15 mV. The solid curves are the result of the fitting by Eq. (\ref{Eq:P(f)}).}\label{fig:HFset}\end{figure}

A schematic illustration of the high-frequency noise measurement setup and the  noise spectra obtained with the setup are  shown in Figs. \ref{fig:HFset}(a) and \ref{fig:HFset}(b), respectively.
The measurement is performed in a dilution refrigerator by utilizing an $LC$ resonator with a peak frequency $f_0$ of 2.55 MHz and a homemade cryogenic amplifier as we reported previously. \cite{Chida2012PRB, Hashisaka2009RSI, Hashisaka2008JPCS}
As shown in Fig. \ref{fig:HFset}(b), the resonant peak at 2.55 MHz is larger for the biased case ($V_{\rm sd}$ = 15 mV) than in the equilibrium case ($V_{\rm sd}$ = 0 mV).
The difference is mainly due to the excess noise.
We evaluate the resonant peak by using the following Lorentzian-like function: \cite{Hashisaka2008PSS, DiCarlo2006RSI}
\begin{equation}
P(f) = P_\text{B} + \frac{P_0}{1+(f^2 - f_0^2)^2 / (f \Delta f)^2}
\label{Eq:P(f)},\end{equation}
where $P_{\rm B}$ is the frequency-independent background noise, $P_{\rm 0}$ is the peak height of the Lorentzian-like function, $f_{\rm 0}$ is the peak frequency, and $\Delta f$ is the full width at half maximum.
By evaluating $\Delta f$ using the equation $\Delta f = (1 / 2 \pi C) [G(V_{\rm sd}) + 1/Z]$, the capacitance $C$ and the impedance $Z$ of the setup are estimated to be 120 pF and 70 k$\Omega$, respectively. \cite{DiCarlo2006RSI, Nishihara2012APL}
The noise spectral density of the device is obtained from the following conventional equation: \cite{Blanter2000PR}
\begin{eqnarray}
P_0(V_{\rm sd}) =  A \biggl( S_V^{\rm out} +&& \left( \frac{Z R}{Z + R} \right)^2 S_I^{\rm out} \nonumber \\
&&+ \left( \frac{Z R}{Z + R} \right)^2 S_{\rm dev}(V_{\rm sd}) \biggr)
\label{Eq:P_0},\end{eqnarray}
where $A$ is the square of the total gain of the amplifier system and $S_V^{\rm out}$ and $S_I^{\rm out}$ are the voltage and current noise generated by the cryogenic amplifier, respectively.
$R = 1/G$ is the resistance of the device and $S_{\rm dev}(V_{\rm sd})$ is the current noise generated at the device, which  consists of  thermal noise $S_{\rm th} = S_{\rm dev}(0)$ and  excess noise $S(V_{\rm sd}) = S_{\rm dev}(V_{\rm sd}) - S_{\rm dev}(0)$.
The setup is calibrated by measuring $S_{\rm th} = 4 k_{\rm B} T G(0)$ at  equilibrium with a quantum point contact on the same device, where $k_{\rm B}$ is the Boltzmann constant.
The obtained parameters of the setup are $A = 3.5 \times 10^5$, $S_V^{\rm out} = 1.0 \times 10^{-19}$ V$^2$/Hz,  and $S_I^{\rm out} = 1 \times 10^{-28}$ A$^2$/Hz, and the base electron temperature of the dilution refrigerator is $T \sim$ 100 mK. \cite{COM:temp}

Because the resonant frequency of the resonator is on the scale of megahertz, which is usually sufficiently high to damp the $1/f$ noise, the resonant peak is expected to be attributed to only the white noise.
However, to ensure that the obtained spectral density is free from any frequency-dependent noise contribution, we need to examine spectral density behavior over a wide frequency range, as discussed in the following text.

\subsubsection{Low-frequency measurements}
\begin{figure}[b]\center \includegraphics[width=1.0\linewidth]{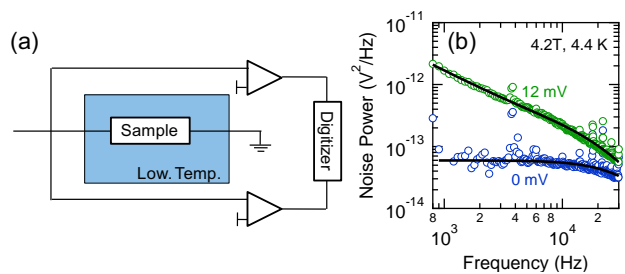}\caption{(Color online) (a) Schematic illustration of the setup for the low-frequency noise measurement utilizing the cross-correlation technique. (b) Noise spectra in the kilohertz range  obtained at $V_{\rm sd}$ = 0  and 12 mV. The solid curves are the result of the fitting by Eq. (\ref{Eq:spec}).}\label{fig:LFset}\end{figure}.

To evaluate a possible contribution of the $1/f$ noise to the excess noise at 2.55 MHz, the noise spectrum in the range from 1  to 100 kHz is obtained with a variable temperature insert (VTI) as in previous experiments. \cite{Hashisaka2008PSS, Sekiguchi2010APL, Arakawa2011APL, Tanaka2012APEX}
The cross-correlation technique is employed as shown in Fig. \ref{fig:LFset}(a) to minimize the external noise from the cables and the amplifiers.
Figure \ref{fig:LFset}(b) shows $P(f)$ obtained in the QHE regime ($V_{\rm sd} = 0$ mV) and the breakdown regime ($V_{\rm sd} = 12$ mV).
 $RC$ damping is obvious for frequencies above 20 kHz.
We have confirmed that the obtained noise above 1 kHz is composed of the  intrinsic noise of the present device caused by the long time averaging in the cross-correlation method.
In the QHE regime, the excess noise in the frequency range from 1  to 20 kHz is frequency independent.
However, the $1/f$ contribution of the noise is observed in the breakdown regime.
Thus, the noise spectrum is expressed by the following equation:
\begin{equation}
P(f) = A \left( S_V^{\rm dev}(V_{\rm sd}) + \frac{a(V_{\rm sd})}{f} \right) \left( \frac{1}{1 + (2 \pi f C R)^2} \right)
\label{Eq:spec},\end{equation}
where $S_V^{\rm dev}(V_{\rm sd}) = R^2 S_{\rm dev}(V_{\rm sd})$ is the contribution of the frequency-independent noise, $a(V_{\rm sd})/f = R^2 S_{1/f}(V_{\rm sd})$ is that of the $1/f$ noise, $1/ [1 + (2 \pi f C R)^2]$ represents the $RC$ damping with a cutoff frequency of $f_c = 1 / 2 \pi C R$, and $C$ is the capacitance between the signal line and the ground through the coaxial cables. 
The curve fitting is performed in the frequency range between 1  and 50 kHz.
The equilibrium noise measurement determines $A$, $C$, and the base electron temperature for the measurement setup as $1.044 \times 10^4$, $400$ pF, and 4.4 K, respectively.

\section{Results and Discussion}
\subsection{Estimation of the noise temperature}
The excess noise at 2.55 MHz is converted to noise temperature $T_{\rm N} = S(V_{\rm sd}) / 4 k_{\rm B} G(V_{\rm sd})$ to study the electron heating accompanied by the QHE breakdown.
In addition, we evaluate the contribution of the $1/f$ noise at 2.55 MHz from the noise spectra obtained from the low-frequency noise measurement to exclude the contribution of unintended low-frequency noise on $S(V_{\rm sd})$ at 2.55 MHz.
In the next sections, the noise temperature in the breakdown regime is discussed with a combination of  high-frequency  and low-frequency noise measurements.

\subsubsection{High-frequency measurements}
\begin{figure}[b]\center \includegraphics[width=1.0\linewidth]{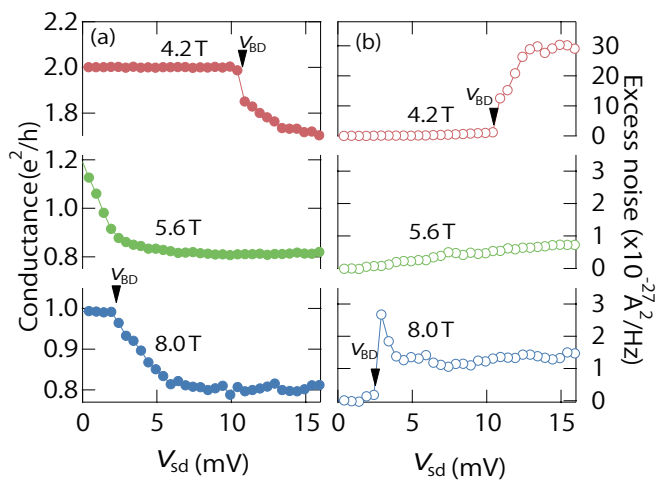}\caption{(Color online) (a) $G$ as a function of $V_{\rm sd}$. (b) $S$ as a function of $V_{\rm sd}$. Measurements are performed at $B = 4.2$, 5.6, and 8.0 T ($\nu = 2.1$, 1.6, and 1.1, respectively) and $T = 100$ mK.}\label{fig:BD}\end{figure}

Figure \ref{fig:BD}(a) shows $G(V_{\rm sd})$ as a function of $V_{\rm sd}$ at $B =$ 4.2 ($\nu = 2.1$), 5.6 ($\nu = 1.6$), and 8.0 T ($\nu = 1.1$).
In the $\nu = 2$ QHE state ($B = 4.2$ T), $G(V_{\rm sd})$ shows quantization as $2e^2/h$ at $V_{\rm sd}$ up to $V_{\rm sd} = 10.4$ mV.
At $V_{\rm sd} = 10.4$ mV $= V_{\rm BD}$, $G(V_{\rm sd})$ abruptly deviates from the quantized value because of the QHE breakdown.
At $V_{\rm sd}$ larger than $V_{\rm BD}$, $G(V_{\rm sd})$ is smaller than $2e^2/h$ owing to the presence of  electron scattering.
The same features are also observed in the $\nu = 1$ QHE state ($B = 8.0$ T).
In this case, $G(V_{\rm sd})$ is quantized as $e^2/h$ and $V_{\rm BD} = 2.0$ mV.
In contrast, in the transition between the two QHE states ($B = 5.6$ T), $G(V_{\rm sd})$ does not exhibit any conductance quantization around $V_{\rm sd} = 0$ mV.

Figure \ref{fig:BD}(b) shows $S(V_{\rm sd})$ as a function of $V_{\rm sd}$, which is simultaneously obtained with $G(V_{\rm sd})$ in Fig. \ref{fig:BD}(a).
In the QHE states ($B = 8.0$ and 4.2 T), $S(V_{\rm sd})$ is strongly suppressed at $V_{\rm sd} < V_{\rm BD}$, reflecting the edge transport of the QHE state with the strong suppression of  backscattering.
At $V_{\rm sd} = V_{\rm BD}$, an abrupt increase in the excess noise is observed; this is  explicit evidence of the transition between the dissipationless state and the dissipative one.

In the transition of the QHE states ($B = 5.6$ T), $ S(V_{\rm sd})$ is very small.
The nominal Fano factor $F = S(V_{\rm sd}) / 2 e V_{\rm sd} G(V_{\rm sd})$ is $5 \times 10^{-4}$, which indicates that our device is sufficiently macroscopic to exclude any shot noise contribution to the excess noise. Note that the device length is much larger than the mean free path of the 2DEG, $l \sim$ 12 $\mu$m.

Now, we assume that the electron heating in the breakdown regime can be regarded as $T_N(V_{\rm sd})$ (the validity of which is discussed in later text).
In the even-integer QHE state, $T_{\rm N}$ is deduced as $\sim \negthickspace10$ K using the typical values of $G(V_{\rm sd})$ and $S(V_{\rm sd})$ in the breakdown regime: $\sim 1.8 e^2/h$ and $30 \times 10^{-27}$ A$^2$/Hz [see Figs. \ref{fig:BD}(a) and 4(b)].
In the odd-integer QHE state, $T_{\rm N}$ is about 1 K for the typical values of $G(V_{\rm sd})$ and $S(V_{\rm sd})$ in the breakdown regime: $\sim 0.8$$e^2/h$ and $1.0 \times 10^{-27}$ A$^2$/Hz [see Figs. \ref{fig:BD}(a) and 4(b)].

We found that the obtained $T_{\rm N}(V_{\rm sd})$ in the breakdown regime is of the order of the energy gap between the Landau levels.
In the odd-integer QHE case, the energy gap is determined by the Zeeman energy $E_{\rm Z} = g \mu_{\rm B} B$, where $g$ is the electron $g$ factor of bulk GaAs and $\mu_{\rm B}$ is the Bohr magneton.
$E_{\rm Z}$ is about 2 K at $B = 8.0$ T, and the obtained $T_{\rm N}$ at $B = 8.0$ T is about 1 K.
In the even-integer QHE case, the energy gap is determined by the cyclotron energy $E_{\rm c} = \hbar \omega_{\rm c}$, where $\hbar$ is the Plank constant and $\omega_{\rm c}$ is the cyclotron angular frequency.
At 4.2 T, $E_{\rm c} \sim 35$ K and $T_{\rm N}$ is deduced as $\sim \negthickspace10$ K.

\begin{figure}[tb]\center \includegraphics[width=.9\linewidth]{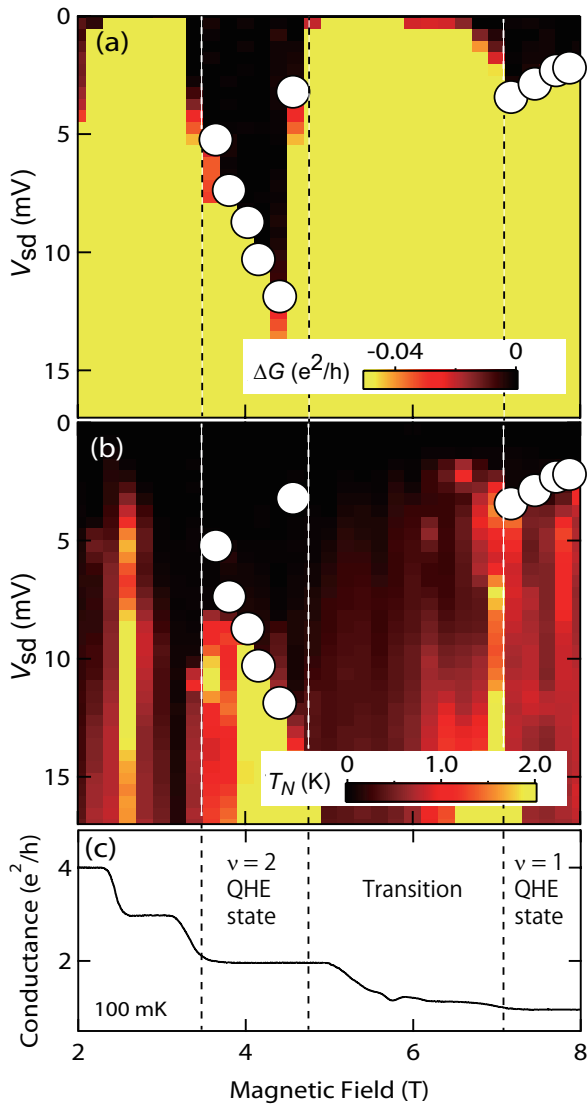}\caption{(Color online) (a) Color plot of $\Delta G = G(V_{\rm sd}) - G(0)$ as a function of $B$ and $V_{\rm sd}$. (b) Color plot of $T_{\rm N}$ as a function of $B$ and $V_{\rm sd}$. (c) Equilibrium conductance as a function of $B$. The open circles in (a) and (b) indicate $V_{\rm BD}$ at each field.}\label{fig:HFdata}\end{figure}

Figure \ref{fig:HFdata}(a) shows a color plot of the conductance deviation from the equilibrium value $\Delta G(V_{\rm sd}) = G(V_{\rm sd}) - G(0)$ as a function of $B$ and $V_{\rm sd}$.
$\Delta G(V_{\rm sd}) = 0$ at a small finite $V_{\rm sd}$ around 4 and 8 T, which reflects the conductance quantization of the QHE state.
In the transition between the QHE states [5--7 T; see Fig. \ref{fig:HFdata}(c)], $G(V_{\rm sd})$ does not have a conductance plateau [$\Delta G(V_{\rm sd}) \neq 0$ at a small $V_{\rm sd}$].
The QHE breakdown is observed as an abrupt collapse of the quantized conductance at finite $V_{\rm sd}$.
The open circles in Figs. \ref{fig:HFdata}(a) and \ref{fig:HFdata}(b) indicate $V_{\rm BD}$ at each field.

Figure \ref{fig:HFdata}(b) shows a color plot of $T_{\rm N}(V_{\rm sd})$ as a function of $B$ and $V_{\rm sd}$.
In the $\nu = 1$ QHE state (around 8 T), because the QHE state is dissipationless, $T_{\rm N}(V_{\rm sd})$ is almost zero around $V_{\rm sd} = 0$ mV.
After the QHE breakdown, at $V_{\rm sd}$ larger than $V_{\rm BD}$, $T_{\rm N}(V_{\rm sd})$ is about 1 K, which falls in the same range as the energy gap between the Landau levels. 
In the transition of the QHE states (around 6 T), the excess noise is rather low owing to the absence of the energy gap.
In the $\nu = 2$ QHE state (around 4 T), the excess noise is strongly (but not perfectly) suppressed when the conductance is quantized.
After the QHE breakdown, $T_{\rm N}(V_{\rm sd})$ increases abruptly to about 10 K.

\subsubsection{Low-frequency measurements}
The above estimation of $T_{\rm N}(V_{\rm sd})$ was based on the assumption that the excess noise is frequency independent.
We validate this assumption with the noise spectrum obtained by using the low-frequency noise measurement.
Unfortunately, the noise spectrum is only obtained at 4.4 K because of our experimental setup.
However, empirically, the $1/f$ noise increases when the temperature increases. Therefore, the $1/f$ noise amplitude at 4.4 K gives us the upper bound of the $1/f$ noise contribution at 100 mK.

\begin{figure}[tb]\center \includegraphics[width=.9\linewidth]{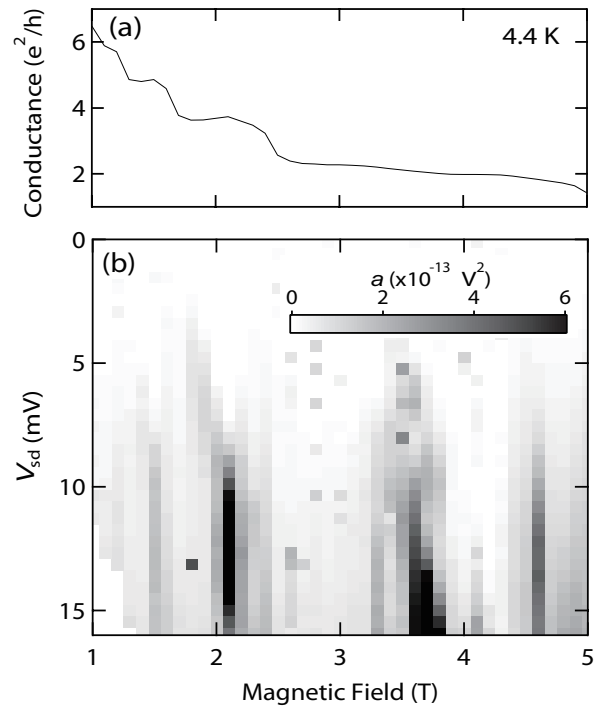}\caption{(a) Equilibrium conductance as a function of $B$ at $T = 4.4$ K. (b) Plot of $a(V_{\rm sd})$ as a function of $V_{\rm sd}$ and $B$.}\label{fig:LFdata}\end{figure}

Figure \ref{fig:LFdata}(a) shows the equilibrium conductance of the device as a function of $B$ obtained with the low-frequency noise measurement setup.
Because the thermal fluctuation at 4.4 K is larger than the Zeeman energy ($E_{\rm z} \sim 2$ K at 8.0 T), the plateau of the $\nu = 1$ QHE state is absent (not shown).
The noise spectrum was well reproduced with Eq. (\ref{Eq:spec}) [see Fig. \ref{fig:LFset}(b)].
Thus, the excess noise is composed of two components: the frequency-independent noise and the $1/f$ noise.

Before evaluating the $1/f$ noise contribution to the excess noise at 2.55 MHz, it is worth considering the frequency-independent component of the excess noise $S(V_{sd}) = S_{\rm dev}( V_{\rm sd}) - S_{\rm dev}(0)$.
In the $\nu = 2$ QHE state (around 4.0 T), $S(V_{sd})$ is strongly suppressed at a $V_{\rm sd}$ smaller than 8 mV, reflecting the presence of dissipationless edge transport of the QHE state.
At $V_{\rm sd} = 10.0$ mV, $S(V_{\rm sd})$ is about $20 \times 10^{-27}$ A$^2$/Hz and it is estimated as $T_{\rm N}(V_{\rm sd}) \sim 7$ K.
The estimated value of $ T_{\rm N}(V_{\rm sd})$ is consistent with the value obtained from the high-frequency noise measurement.

Figure \ref{fig:LFdata}(b) shows a plot of the $1/f$ noise amplitude $a(V_{\rm sd})$ as a function of $V_{\rm sd}$ and $B$.
The maximum value of $a(V_{\rm sd})$ is observed at $B = 3.6$ T and $V_{\rm sd} = 15$ mV and is about $6 \times 10^{-13}$ V$^2$.
From the maximum value of $a(V_{\rm sd})$, the $1/f$ noise amplitude at 2.55 MHz is estimated as $\negthickspace1.2 \times 10^{-27}$ A$^2$/Hz by using the empirical relation $S_{1/f}(f) = G(V_{\rm sd})^2  a/ f $, which is less than a few percent of the excess noise in the breakdown regime at the $\nu = 2$ QHE state ($\sim \negthickspace30 \times 10^{-27}$ A$^2$/Hz).
Because the $1/f$ noise is empirically reduced by the temperature decrease, the estimation of the noise temperature from the excess noise is justified.

\subsection{Observation of the precursor phenomenon of the QHE breakdown}
In this section, we report the finite excess noise at a $V_{\rm sd}$ smaller than $V_{\rm BD}$. 
The excess noise prior to the breakdown of the QHE behaves in a way closely related to the prebreakdown of the QHE. \cite{Ebert1983JPC, Komiyama1985SSC, Meziani2004JAP}

\begin{figure}[b]\center \includegraphics[width=1.0\linewidth]{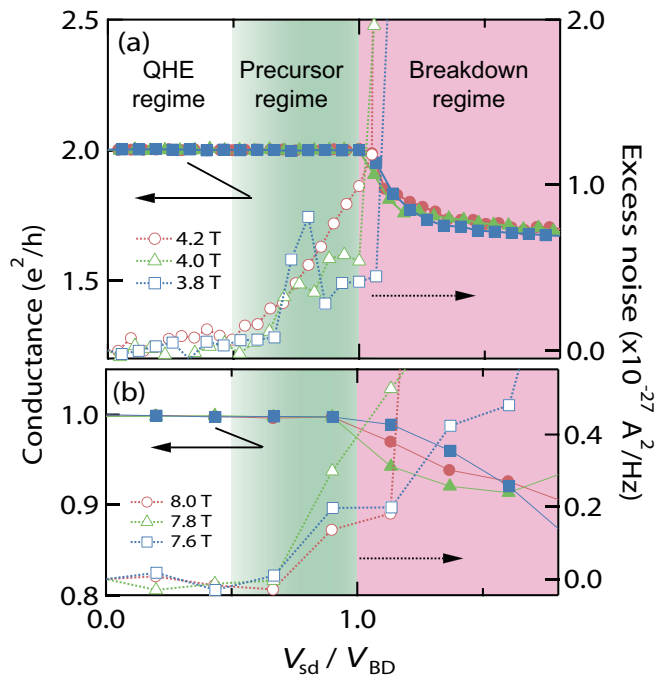}\caption{(Color online) $G(V_{\rm sd})$ and $S(V_{\rm sd})$ as a function of $V_{\rm sd}$ obtained at (a) $\nu \sim 2$ and (b) $\nu \sim 1$. The range of $V_{\rm sd}$ is divided into a QHE regime, a precursor regime, and a breakdown regime.}\label{fig:Prec}\end{figure}

Figure \ref{fig:Prec}(a) shows $G(V_{\rm sd})$ and $S(V_{\rm sd})$ as a function of the normalized source-drain bias voltage ($V_{\rm sd} / V_{\rm BD}$) obtained around the $\nu = 2$ QHE state.
Data obtained at $B = 4.2,$ 4.0, and 3.8 T are plotted as circles, triangles, and squares, respectively.
Those plots have almost the same $V_{\rm sd}$ dependence, which is described as follows:
At $V_{\rm sd} / V_{\rm BD} = 1$, $G(V_{\rm sd})$ deviates from $2e^2/h$ and $S(V_{\rm sd})$ increases abruptly owing to the QHE breakdown.
An important characteristic of $S(V_{\rm sd})$ is the finite excess noise at $V_{\rm sd} < V_{\rm BD}$.
Typically, the excess noise reaches about $0.6 \times 10^{-27}$ A$^2$/Hz at $V_{\rm sd} / V_{\rm BD} = 0.95$.
This is about an order smaller than the value reached in the breakdown regime.

The finite excess noise is also observed in the vicinity of the $\nu = 1$ QHE state.  Figure \ref{fig:Prec}(b) shows data obtained at $B = 8.0,$ 7.8, and 7.6 T.
In spite of the low density of data points, finite excess noise is always observed at $V_{\rm sd}$ smaller than $V_{\rm BD}$. A typical value of $S(V_{\rm sd})$ is about $0.2 \times 10^{-27}$ A$^2$/Hz at $V_{\rm sd} / V_{\rm BD} = 0.90$.

To clarify the following discussion, we divide $V_{\rm sd}$ into three regions: the QHE regime, the precursor regime, and the breakdown regime, as shown in Fig. \ref{fig:Prec}.
In the QHE regime (typically, at $V_{\rm sd}/V_{\rm BD} \lesssim 0.5$), $G(V_{\rm sd})$ is quantized and $S(V_{\rm sd})$ is strongly suppressed.
In the precursor regime, although $G(V_{\rm sd})$ is still quantized,  finite $S(V_{\rm sd})$ is observed even when $V_{\rm sd}$ is smaller than $V_{\rm BD}$.
At $V_{\rm sd}/V_{\rm BD} > 1$, the QHE breakdown causes  $G(V_{\rm sd})$ to deviate from the quantized value.

The $V_{\rm sd}$ dependence of $S(V_{\rm sd})$ is different in the three $V_{\rm sd}$ regions.
In the QHE regime, even though $S(V_{\rm sd})$ is strongly suppressed, a small, finite $S(V_{\rm sd})$ of less than $10^{-28}$ A$^2$/Hz is observed.
$S(V_{\rm sd})$ shows a clear quadratic $V_{\rm sd}$ dependence.
Hence, we conclude  that $S(V_{\rm sd})$ in the QHE regime originates from Joule heating at the Ohmic contacts.
The quadratic $V_{\rm sd}$ dependence deviates in the precursor regime because of additional noise.
The emergence of this additional noise is not transitional but appears as a ``crossover."
Therefore, the boundary between the QHE regime and the precursor regime is not clear.
In contrast, the boundary between the precursor state and the breakdown state is obvious, as the QHE breakdown is a transition.
In the breakdown regime, $S(V_{\rm sd})$ is almost constant, as seen in Fig. \ref{fig:BD}.

Because the additional excess noise in the precursor regime is universally observed in different Landau fillings (see Fig. \ref{fig:Prec}), the excess noise is related to the universal behavior of the QHE regime.

\subsection{Possible origins of the precursor phenomenon}
\begin{figure}[tb]\center \includegraphics[width=.9\linewidth]{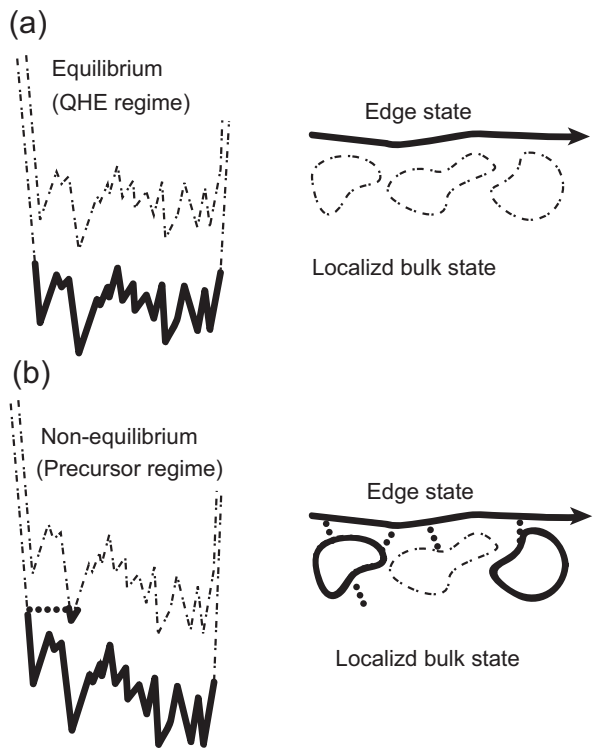}\caption{Schematic illustration of the edge states and the localized bulk states in (a) the QHE regime and (b) the precursor regime. (left) Energy-position plot; the vertical axis is the energy and the horizontal axis is the position across the counterflowing edge channels. (right) Real space image of the edge and the bulk states. The dashed curve and the solid curve represent unoccupied states and occupied states, respectively. The dotted line denotes electron tunneling between the edge state and the bulk states.}\label{fig:Mecha}\end{figure}

Let us start by discussing characteristics of electron transport in the QHE regime.
In this regime, electrons flow in the conductive edge channels.
Counterflowing channels are spatially separated by the bulk insulating state, as Halperin demonstrated. \cite{Halperin1982PRB}
Hence, backscattering of electrons is strongly suppressed.
In the bulk, electrons and holes are localized in puddles with a typical size on the order of 100 nm. \cite{Finkelstein2000Sci, Zhitenev2000Nat}
This conductive edge and insulating bulk picture is schematically shown in Fig. \ref{fig:Mecha}(a).
The Fermi surface is placed at the zero-gap metallic state at the edge and at the energy gap in the bulk.
This energy gap protects the electrons being transported by backscattering. \cite{Buttiker1988PRB}

The edge transport of the QHE state breaks down according to the following scenario.
First, electrons in the edge state tunnel to the localized bulk state.
The excited electrons are accelerated by the Hall electric field.
When the accelerated electrons obtain sufficient energy, the avalanche type electron scattering \cite{Komiyama1996PRL} and/or the quasi inter-Landau-level scattering (QUILLS) assisted by acoustic phonons \cite{Chaubet1998PRB} cause the electron heating. 

From this scenario of the QHE breakdown, the presence of electron tunneling without the avalanche and/or the QUILLS is reasonable and we believe that this is what we observe in the precursor regime.
A schematic image of the QHE state is shown in Fig. \ref{fig:Mecha}(b).
In this $V_{\rm sd}$ region, electrons  tunnel back and forth between the edge states and the localized puddles.
Such tunneling effects have been studied extensively and are plausible \cite{Machida2001PRB, Peled2003PRL, Couturaud2009PRB}.
When $V_{\rm sd}$ becomes larger than $V_{\rm BD}$, the electrons in the bulk generate the avalanche and/or the QUILLS, and the QHE state breaks down.

The origin of the excess noise in the precursor regime is not simply a result of the electron tunneling itself as the excess noise is frequency independent.
The origin of the excess noise is likely to be caused by the increase in the effective electron temperature.
The injection of  hot electrons into the localized bulk state through  electron tunneling increases the effective electron temperature inside the bulk state, and  electron tunneling between the edge state and the bulk state results in the broadening of the energy distribution inside the edge state.
The broadening of the energy distribution is observed as a finite increase in $S(V_{\rm sd})$ in the voltage-biased QHE state.
\\

\section{Summary}
We performed noise measurements for a device in a nonequilibrium quantum Hall effect state and identified two distinct components of excess noise.
The first one  originates from avalanche-type electron heating of the QHE breakdown.
The other is most likely to be related to the prebreakdown of the QHE, which clearly indicates the finite dissipation prior to the breakdown of the QHE.
As the noise measurement shows such a high sensitivity to dissipation of the current, further experimental effort using the noise measurement in the QHE state would clarify in more detail how the QHE state breaks down. 

\section*{Acknowledgments}
We  appreciate fruitful discussions with M. Hashisaka, Y. Yamauchi, S. Nakamura, Y. Tokura, K. Muraki, T. Fujisawa, K. Oto, and M. Kawamura. This work is partially supported by the JSPS Funding Program for Next Generation World-Leading Researchers and a Grant-in-Aid for JSPS Fellows.


\begin{thebibliography}{99}
\bibitem{COM:temp} This temperature results from the heat current through the signal line for the resistive-detection NMR. Without the signal line, the base electron temperature of the setup is below 20 mK.
\end{thebibliography}
\end{document}